\documentclass[12pt,english,floatfix,superscriptaddress,aps,preprint,showpacs]{revtex4}
\usepackage{amsmath}
\usepackage{amssymb}
\usepackage{amsbsy}
\usepackage{amsfonts}
\usepackage{amsopn}
\usepackage{amstext}
\usepackage{graphicx}
\usepackage{amssymb}
\usepackage{amsfonts}
\usepackage{amsmath}
\usepackage{graphicx}
\usepackage[english]{babel}
\usepackage{color}
\usepackage[dvips]{epsfig}
\usepackage[dvips]{graphicx}
\usepackage{float}
\usepackage{units}
\usepackage{textcomp}
\usepackage{babel}
\DeclareMathOperator{\sech}{sech}
\begin{document}

\title{Gravity localization in sine-Gordon braneworlds}
\author{W. T. Cruz}
\email{wilamicruz@gmail.com}
\affiliation{Instituto Federal de Educa\c{c}\~{a}o, Ci\^{e}ncia e Tecnologia do Cear\'{a} (IFCE), Campus Juazeiro do Norte - 63040-540 Juazeiro do Norte-Cear\'{a}-Brazil}
\author{R. V. Maluf}
\email{r.v.maluf@fisica.ufc.br}
\affiliation{Departamento de F\'{i}sica - Universidade Federal do Cear\'{a} (UFC) - C.P. 6030, 60455-760
Fortaleza-Cear\'{a}-Brazil}
\author{L. J. S. Sousa}
\email{luisjose@fisica.ufc.br}
\affiliation{Instituto Federal de Educa\c{c}\~{a}o, Ci\^{e}ncia e Tecnologia do Cear\'{a} (IFCE), Campus Canind\'{e} - 62700-000 Canind\'{e}-Cear\'{a}-Brazil}
\author{C. A. S. Almeida}
\email{carlos@fisica.ufc.br}
\affiliation{Departamento de F\'{i}sica - Universidade Federal do Cear\'{a} (UFC) - C.P. 6030, 60455-760
Fortaleza-Cear\'{a}-Brazil}
\date{\today}

\begin{abstract}
In this work we study two types of five-dimensional braneworld models given by sine-Gordon potentials. In both scenarios, the thick brane is generated by a real scalar field coupled to gravity. We focus our investigation on the localization of graviton field and the behaviour of the massive spectrum. In particular, we analyse the localization of massive modes by means of a relative probability method in a Quantum Mechanics context. Initially, considering a scalar field sine-Gordon potential, we find a localized state to the graviton at zero mode. However, when we consider a double sine-Gordon potential, the brane structure is changed allowing the existence of  massive resonant states. The new results show how the existence of an internal structure can aid in the emergence of  massive resonant modes on the brane.
\end{abstract}


\maketitle

\section{Introduction\label{intro}}

The interest in thick brane models has grown in the context of five-dimensional warped spacetime due to its advantages over the Randal-Sundrum (RS) scenarios \cite{RS}. Several authors have noticed the presence of resonances while analysing gravity localization in RS models \cite{lifetime, csaki1, csaki2, metastable, adriana}. The same structures have also been detected in the study of gravitons in five-dimensional branes generated dynamically by topological defects \cite{novochineses, nosso8, nosso4, creu,novochineses2}.
If we are interested in braneworlds with one extra dimension, we can consider models that support kink solutions. In this case, scalar fields coupled to gravity in the presence of potentials like $\lambda \phi^4$ can provide kink solutions, which are responsible for the appearance of new and interesting thick braneworld scenario. Alternatively, bounce-like solutions can also be obtained by sine-Gordon (SG) potentials. Therefore, some works have studied such solutions as brane models \cite{chineses_sine, bazeia_sine1, bazeia_sine2}. However, a complete study of the graviton massive spectrum in defects generated by sine-Gordon potentials are still absent.

Thick braneworld can be further generated by two-kink solutions with the advantage that these solutions have internal structure that may interfere with the field localization mechanisms. Such defects can be obtained from a deformation of a $\lambda \phi^4$ potential \cite{f1,f2,f3,aplications,brane} and seem to be composed of two standard kinks. An example of this is the Block brane model, on which the study  of the field localization has been carried out by some of us in Refs. \cite{nosso1,nosso3,nosso5,nosso6,nosso7}. An alternative way to obtain two-kink solutions is consider a double sine-Gordon (DSG) potential, which has been investigated for instance in Refs. \cite{aplications,brane}. In this type of model, it is not necessary to take a deformation procedure in order to achieve the brane with an internal structure. 

The DSG  model has applications in several physical contexts, for example, it was applied to the analysis of magnetic solitons in superfluid $^3$He \cite{dsg_ap1}, study of thermodynamical properties in the magnetic chains \cite{dsg_ap2} and  polymers \cite{dsg_ap3}. A perturbation theory for the DSG equation is presented in \cite{dsg3}, and the author in Ref. \cite{dsg4} has showed two different methods in order to solve this equation. In particular, the study of velocity resonances in the classical scattering of DSG kinks was addressed in \cite{dsg5}.

In this work, the SG and the DSG potentials are adopted to construct  braneworld models where we analyse the localization of gravitons with focus on the behaviour of the massive spectrum. The graviton equations of motion are converted into a Schr\"{o}dinger-like equation and we search for resonances on the resulting wave functions. In the context of quantum mechanics, the normalized wave functions are used to evaluate by numerical methods the probability to find a massive mode in terms of the position along the extra dimension. We are specifically interested in the modes with high probability of being located at the brane position.

This work is organized as follows. In the second section, we review the brane setup obtained by the sine-Gordon potential. In the next section, we introduce the DSG model and the braneworld corresponding. We addressed the graviton massive spectrum and search for resonances on the SG and DSG models in the fourth section. Finally, we discuss our results and present our conclusions in the last section.

\section{Sine-Gordon brane \label{sec:sine}}
\begin{figure*}
 \centering
    \includegraphics[width=0.95\textwidth]{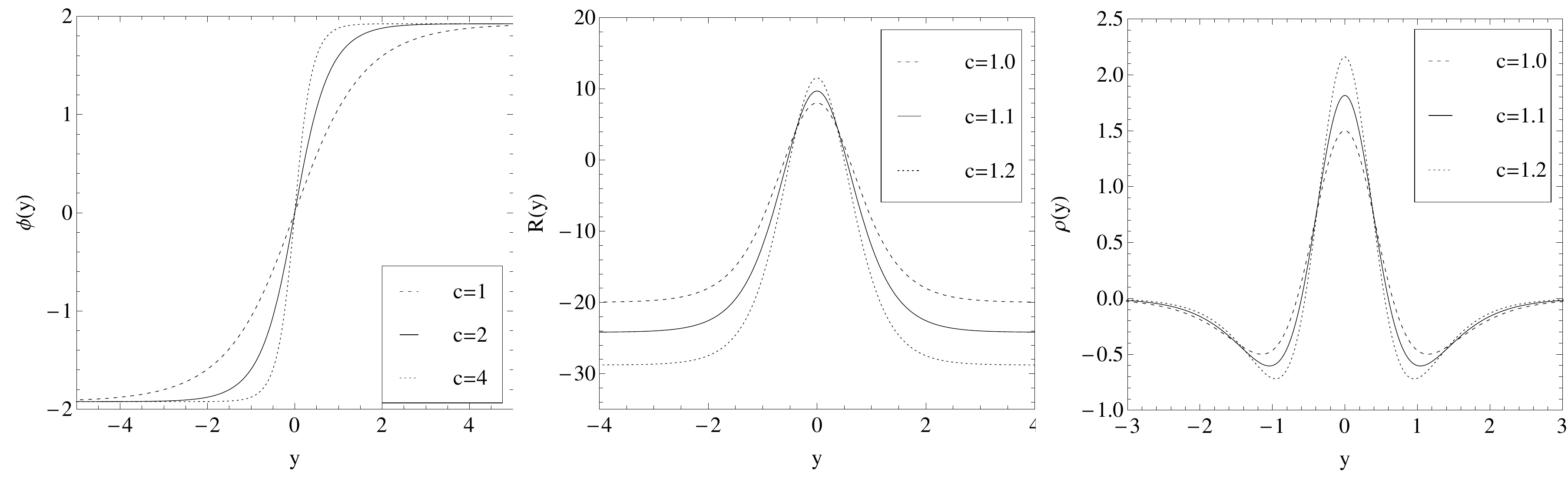}
 \caption{The kink solution (left), curvature scalar (center) and the energy density (right). We have used $b=1$.}
 \label{fig:sine0}
\end{figure*}

Let us now focus on the thick brane scenario constructed by a real scalar field $\phi$ coupled to gravity. Such scalar field depends only on the extra dimension $y$. The spacetime is an AdS $D=5$ with metric $ds^{2}=e^{2A(y)}\eta_{\mu\nu}dx^{\mu}dx^{\nu}+dy^{2}$. The Minkowski spacetime metric is $\eta_{\mu\nu}$ with signature $(-,+,+,+)$ and the indices $\mu$, $\nu$ run over $0$ to $3$. Note that the warp factor is written in terms of the function $A(y)$ that will be defined by an appropriate choice of the potential. Considering the action
\begin{equation}\label{eq:action}
S=\int d^{5}x\sqrt{-G}\Big[\frac{1}{4}R-\frac{1}{2}(\partial\phi)^{2}-V(\phi)\Bigr],
\end{equation} where $R$ is the scalar curvature, the equations of motion take the general form
\begin{equation}
\phi'^{2}-2V(\phi)  =  6A'^{2}\label{eq:eqmov1}
\end{equation}
\begin{equation}
\phi'^{2}+2V(\phi)  =  -6A'^{2}-3A''\label{eq:eqmov2}
\end{equation}
\begin{equation}
\phi''+4A^{\prime}\phi'  =  \frac{\partial V}{\partial \phi},
\label{eq:eqmov3}
\end{equation}
where prime stands for derivative with respect to $y$. Considering the potential
\begin{equation} V(\phi)=\frac{1}{8}\left(\frac{\partial W}{\partial\phi}\right)^{2}-\frac{1}{3}W^{2},
\end{equation}
one obtains the first-order equations
\begin{equation}\label{eq:firstorder}\phi^{\prime}=\frac{1}{2}\frac{\partial W}{\partial\phi},
\end{equation}
\begin{equation} A^{\prime}=-\frac{1}{3}W.
\label{eq:A}
\end{equation}
The above equations were obtained by a method widely used in the literature, which consists of writing the potential $V(\phi)$ in terms of the superpotential $W=W(\phi)$, a smooth function of the field $\phi$ \cite{bazeia1,bazeia2,bazeia3,bazeia4,shif,alonso,de,cvetic,skenderis}. Thus, the sine-Gordon potential is obtained by choosing the following superpotential \cite{gremm}
\begin{equation}\label{eq:supot}
W(\phi)=3 b c \sin\left(\sqrt{\frac{2}{3 b}\phi}\right),
\end{equation}
which provides the potential
\begin{equation}\label{eq:sgpot}
V(\phi)= \frac{3}{8}b c^2 \left[1- 4 b+ \left(1+ 4b\right) \cos\left(\sqrt{\frac{8}{3b}}\phi\right)\right].
\end{equation}
Finally, considering the superpotential (\ref{eq:supot}), the first order equation (\ref{eq:firstorder}) yields the solution
\begin{equation}
\phi(y)=\sqrt{6b}\arctan\lbrace\tanh\left[\frac{1}{6}\left(3 c y+ \sqrt{\frac{6}{b}}C_1\right)\right]\rbrace \label{phi},
\end{equation}
which is plotted in Fig. (\ref{fig:sine0}). Integrating twice the sum of equations (\ref{eq:eqmov1}) and (\ref{eq:eqmov2}) give us the warp factor 
\begin{equation}\label{eq:warpfactor}
e^{2A(y)}= {\cosh\left[c y + \frac{\sqrt{\frac{2}{3}} C_
1}{\sqrt{b}}\right]}^{-2 b}.
\end{equation}

To center the defect at $y = 0$, we set $C_1 = 0$ henceforward. The constant $c$ controls the thickness of the defect and $b$ defines the value of $\phi(y)$ when $y \rightarrow\pm\infty$.

Hereinafter the Ricci scalar assumes the form
\begin{equation}
R= 8 b c^2 \sech(c y)^2 - 20 b^2 c^2 \tanh(c y)^2.
\end{equation}
In the brane localization, the curvature scalar assumes a positive value. However, far from the brane, $R$ tends to a negative constant, characterizing the $AdS_{5}$ limit for the bulk.  The presence of regions with positive Ricci scalar could in principle be connected with the possibility that some massive states may be quasi-localized on the brane as we will investigate in the following sections.

Since the expressions of $\phi(y)$, $A(y)$ and $V(\phi(y))$ are determined, we can write the matter energy density along the extra dimension, given by $e^{2A}(1/2\phi'^2+V)$, in the form expressed as
\begin{eqnarray}
\rho(y)=\frac{3}{8} b c^2 \cosh(c y)^{-2 b}\big[1 - 4 b + (1 + 4 b)\times \\\nonumber \cos\big(4 \arctan(\tanh(\frac{c y}{2}))\big) + 
   2 \sech(c y)^2\big].
\end{eqnarray}
Observing the solutions for $\rho(y)$ and $R(y)$ in Fig. (\ref{fig:sine0}) we note that increasing the values of the constant $c$ makes the brane thinner.

\section{Double sine-Gordon brane \label{sec:dsg}}
In this section, we construct a new thick brane scenario from an extension of the basic SG model. We start again from the action of a scalar field coupled to gravity
\begin{equation}\label{eq:action_dsg}
S=\int d^{5}x\sqrt{-G}\Big[\frac{1}{4}R-\frac{1}{2}(\partial\Phi)^{2}-\mathcal{V}(\Phi)\Bigr].
\end{equation}
In flat space-time the so-called double sine-Gordon model can be obtained by the following potential \cite{analytical_dsg}:
\begin{equation}
\overline{V}(\Phi)= 1 -\cos(2\Phi) + a\big(1 - \cos(\Phi)\big).
\end{equation}
\begin{figure}[h]
 \centering
    \includegraphics[width=0.5\textwidth]{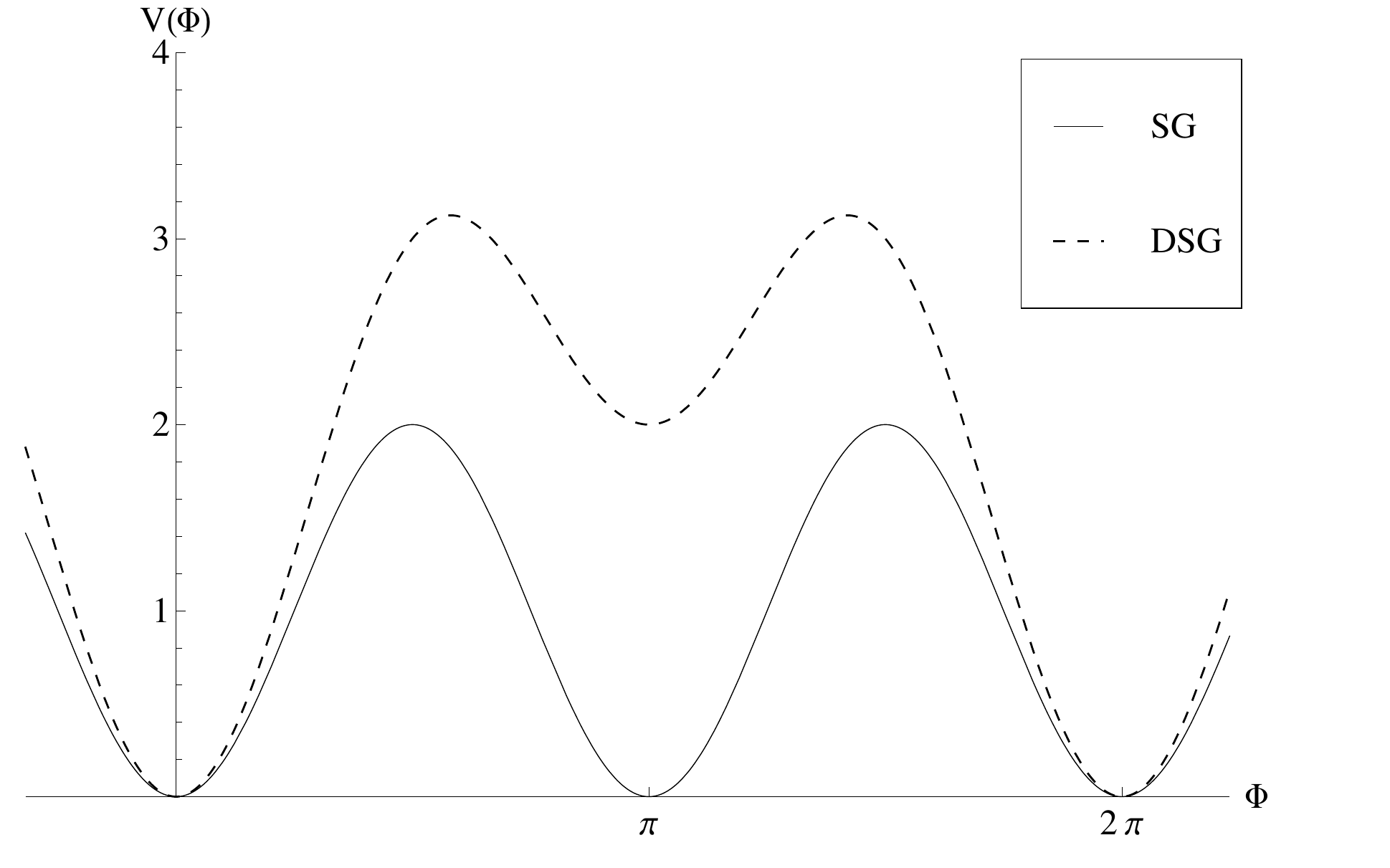}
 \caption{Sine-Gordon and double sine-Gordon potentials in flat space-time.}
 \label{fig:sine1}
\end{figure}
As showed in the second section, the SG potential (\ref{eq:sgpot}) provides us kink solutions that interpolate between the maxima of the potential. In Fig.  (\ref{fig:sine1}) we compare the SG and DSG potentials. We note that in the DSG potential the absolute minimum at $\Phi=0$ persists, and we have the raising of a relative minima at $\Phi=\pi$. As showed in previous works \cite{dsg3,dsg4,dsg1,dsg2}, the changing in the original vacua of the SG model modifies the scalar field solution. On the DSG potential in flat space-time, the kink solution interpolates between the two vacua ($ 0$ and $2\pi$) with a transient state at $\pi$. In gravitational scenario, this feature reveal us new peculiarities of the brane geometry.

Next, we write the potential $\overline{V}(\phi)$ in the form
\begin{equation}
\overline{V}(\Phi)=\frac{1}{8}\left(\frac{d\mathcal{W}}{d\Phi}\right)^2,
\end{equation}
and so we can determine another superpotential such that
\begin{eqnarray}\label{eq:wdsg}
\mathcal{W}(\Phi)= 4 \cos(\Phi/2) \sqrt{a + 2\big(1+ \cos(\Phi)\big)} + \\\nonumber 2 a \ln\left[2 \cos(\Phi/2) + \sqrt{a + 2\big(1+ \cos(\Phi)\big)}\right].
\end{eqnarray}

Returning to the curved space-time, the equations of motion resulting from the action  (\ref{eq:action_dsg}) give us first-order equations if the potential is written in terms of the superpotential as
\begin{equation}
\mathcal{V}(\Phi)=\frac{1}{8}\left(\frac{\partial \mathcal{W}}{\partial\Phi}\right)^{2}-\frac{1}{3}\mathcal{W}^{2}.
\end{equation}
Thus, this prescription allows us to obtain the first-order equations as
\begin{equation}\label{eq:firstorder-dsg}\Phi^{\prime}=\frac{1}{2}\frac{\partial \mathcal{W}}{\partial\Phi},
\end{equation}
\begin{equation} \mathcal{A}^{\prime}=-\frac{1}{3}\mathcal{W}.
\label{eq:A_dsg}
\end{equation}
In terms of the superpontential $\mathcal{W}(\Phi)$ defined in Eq. \eqref{eq:wdsg}, we get the new DSG potential in curved space-time, given by

\begin{eqnarray}
\mathcal{V}(\Phi)= 1 -\cos(2\Phi) + a\big(1 - \cos(\Phi)\big) - \frac{1}{3}\Bigg[4 \cos(\Phi/2) \sqrt{a + 2\big(1+ \cos(\Phi)\big)} + \\ \nonumber
   2 a \ln\left(2 \cos(\Phi/2) + \sqrt{a + 2\big(1+ \cos(\Phi)\big)}\right)\Bigg]^2.
\end{eqnarray}

\begin{figure}[h]
 \centering
    \includegraphics[width=0.4\textwidth]{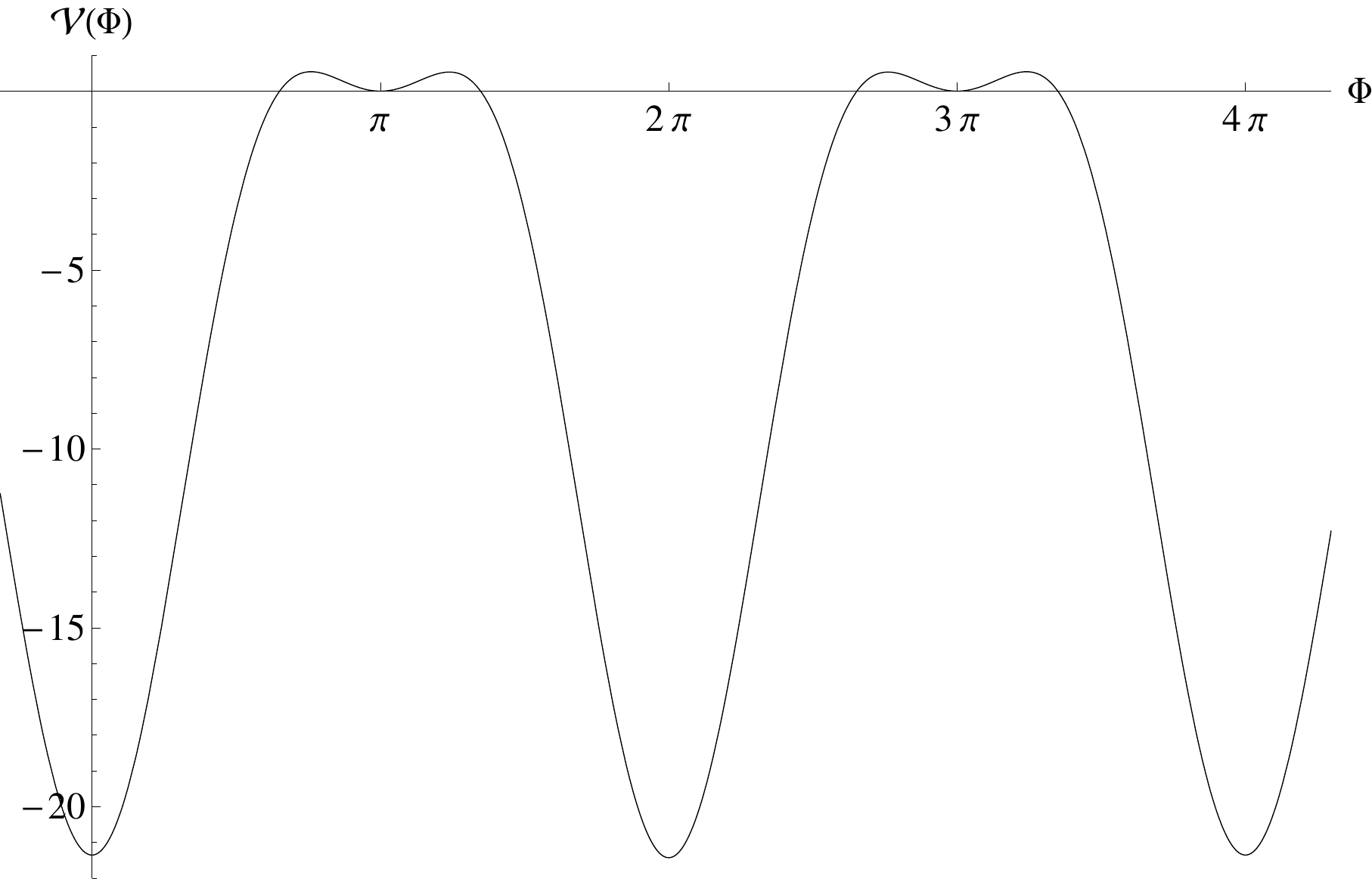}
 \caption{Double sine-Gordon potential in gravitational scenario.}
 \label{fig:sine2}
\end{figure}

In Fig. (\ref{fig:sine2}), we observe that the DSG potential obtained keeps similar behaviour that in the case of the absence of gravity. It is expected that the solution to the scalar field exhibits a transient state among the vacua due to the raising of  intermediate minima in $\mathcal{V}(\Phi)$.

\begin{figure*}
 \centering
    \includegraphics[width=0.95\textwidth]{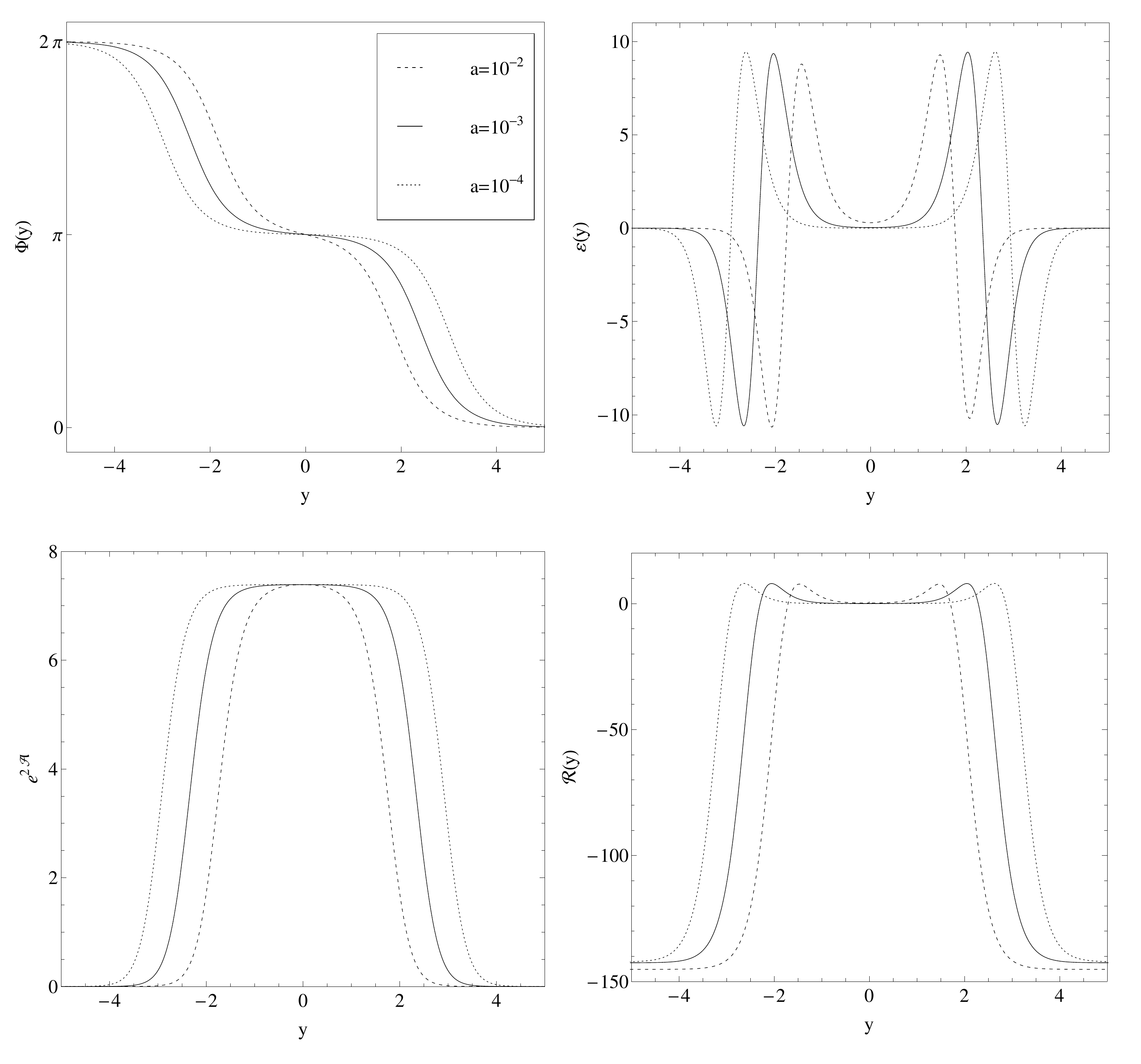}
 \caption{The kink solution (top left), energy density (top right), warp factor (bottom left) and curvature scalar (bottom right).}
 \label{fig:sine3}
\end{figure*}

The new solution  to the scalar field on the DSG scenario must be obtained from the first-order equation (\ref{eq:firstorder-dsg}). Due to the more complex structure of $\mathcal{W}(\Phi)$ than in the SG model we are unable to evaluate $\Phi$ analytically. Thus, we plot the numerical solution for $\Phi$ in Fig. (\ref{fig:sine3}). This new kink solution interpolates between the vacua at $0$ and $2\pi$ with a transient region related to the intermediate minimum of the DSG potential at $\pi$. Similar structures, called two-kink solutions, were obtained in previous works from a deformation procedure of $\Phi^4$ potentials \cite{aplications, brane}. 

The solution for the function $\mathcal{A}(y)$ is obtained from equation
\begin{equation}\label{eq:eqA}
\mathcal{A}''(y)=-2/3\Phi'(y)^2,
\end{equation} that is the sum of the equations (\ref{eq:eqmov1}) and (\ref{eq:eqmov2}). Based on numerical data for $\Phi$, we construct the warp factor, that is plotted in Fig. (\ref{fig:sine3}). In the same way, the energy density
\begin{equation}
\varepsilon(y)=e^{2\mathcal{A}}\left(\frac{1}{2}\Phi'^2+\mathcal{V}\right),
\end{equation}
and the curvature scalar
\begin{equation}
\mathcal{R}(y)=-(8\mathcal{A}''+20\mathcal{A}'^2),
\end{equation}
are also plotted in Fig. (\ref{fig:sine3}).

The DSG model provides new features about the core of the brane that are not present in the SG model. We note the raising of a flat region at the brane location and also a splitting on the matter energy density. The appearance of two new maxima in $\mathcal{R}(y)$ and $\varepsilon$ was interpreted in Refs. \cite{dionisio,brane} as the emergence of a brane with an internal structure. When the constant $a$ is reduced, the two kinks that compose the structure of $\Phi$ separate themselves. The same aspect is observed on the maxima of the energy density and curvature scalar, which are related with the position of each kink in $\Phi$.

\section{Massive spectrum and resonances}\label{sec:massive}

We now focus our attention on the gravity localization in SG and DSG braneworlds. The main objective is to search for resonant states on the massive spectrum. For this goal, we must obtain a Schr\"{o}dinger-like equation to the graviton on the fifth dimension. Initially, we perform a tensor perturbation of the metric as
\begin{equation}
ds^2=e^{2A(y)}(\eta_{\mu\nu}+\epsilon h_{\mu\nu})dx^\mu
dx^\nu+dy^2,
\end{equation}
where $h_{\mu\nu}=h_{\mu\nu}(x,y)$ satisfies the transverse-traceless (TT) conditions and represents the graviton. So when we consider only the TT tensor mode to the metric fluctuations, namely $\overline{h}_{\mu\nu}$, its equations of motion take the simplified form \cite{gremm, kehagias, de}:
\begin{equation}\label{eq:grav}
\overline{h}_{\mu\nu}^{\prime\prime}+4A^{\prime}
\overline{h}_{\mu\nu}^{\prime}=e^{-2A}\partial^2\overline{h}_{\mu\nu}^{\prime},
\end{equation}
where $\partial^2$  is the four-dimensional wave operator. Introducing the transformation $dz=e^{-A(y)}dy$ and choosing an ansatz containing a bulk wave function times a space plane wave, $\overline{h}_{\mu\nu}(x,z)=e^{ip\cdot
x}e^{-\frac{3}{2}A(z)}U_{\mu\nu}(z)$, we can rewrite the Eq. \eqref{eq:grav} as a
Schr\"{o}dinger-like equation given by
\begin{equation}\label{eq:schro}
-\frac{d^2U(z)}{dz^2}+V(z)\,U(z)=m^2\,U(z),
\end{equation}
with the potential $V(z)=\frac32\,A^{\prime\prime}(z)+\frac94\,A^{\prime2}(z)$. Note that we have omitted the space-time indices and introduced the mass parameter such that $p^{2}=-m^{2}$. The Eq. (\ref{eq:schro}) does not lead to tachyonic states and one has a normalizable zero mode solution as shown in Ref. \cite{dionisio}.

The coupling of the massive modes with the matter on the brane is established in terms of the solution to the Schr\"{o}dinger-like equation at $z=0$. The solutions to $U(z)$ acquire plane wave structure when $m^2\gg V_{max}$ because the potential  represents only a small perturbation. Thus, if there are resonant modes we expect that they must emerge with $m^2\leq V_{max}$.

In general, we can define a resonant mode as solutions to $U(0)$ with large amplitudes inside the brane in comparison with its values far from the defect. To seek for such structures we adopt a largely used method based on the relative probability $N(m)$ \cite{chineses1,chineses2,chineses3, chineses4, chineses5} given by
\begin{equation}\label{rel_prob}
N(m)=\frac{\int_{-z_{b}}^{+z_{b}}|U_{m}(z)|^2dz}{\int_{-z_{max}}^{+z_{max}}|U_{m}(z)|^2dz}.
\end{equation}
From Eq. (\ref{eq:schro}) we can consider $\zeta|U_{m}(z)|^2$  as the probability for finding the mode at the position $z$, where $\zeta$ is a normalization constant. Thus, resonant modes will be identified by peaks in $N(m)$.

To evaluate the relative probability, we take a narrow integration range around the brane $-z_b<z<z_b$ with $z_b=0.1z_{max}$ \cite{chineses3}, and the massive modes considered inside a box with borders $|z|=z_{max}$ far from the turning points of the potential \cite{chineses1, chineses4,ca}. It is known that the choice of the integration interval $z_b=0.1z_{max}$ does not interfere with the values of the masses to the possible resonance modes \cite{nosso8}. Since we have a symmetric brane, we use $U(0)=1$ and $U'(0)=0$.

\subsection{SG brane}
Let us now consider the warp factor in the SG scenario. Starting from the solution to $A(y)$ given by Eq. (\ref{eq:warpfactor}), and setting $b=1$ for simplicity, we take the well-known transformation $dz= e^{-A(y)}dy$ so that the warp function becomes
\begin{equation}
A(z)=-\ln(\sqrt{z^2c^2+1}).
\end{equation}
As a matter of fact, there is a graviton zero mode localized solution. For $m=0$, the equation (\ref{eq:schro}) admits the solution $U(z)=1/{(1+c^2 z^2)}^{3/4}$. In Ref. \cite{gremm}, the author has noticed the existence of a zero mode localized to the graviton but have excluded the existence of resonances on the massive spectrum without having evaluated $U_0(m)$.  In fact, when such work was published, the methods to detect resonant states have not been developed yet.

\begin{figure*}
 \centering
    \includegraphics[width=0.95\textwidth]{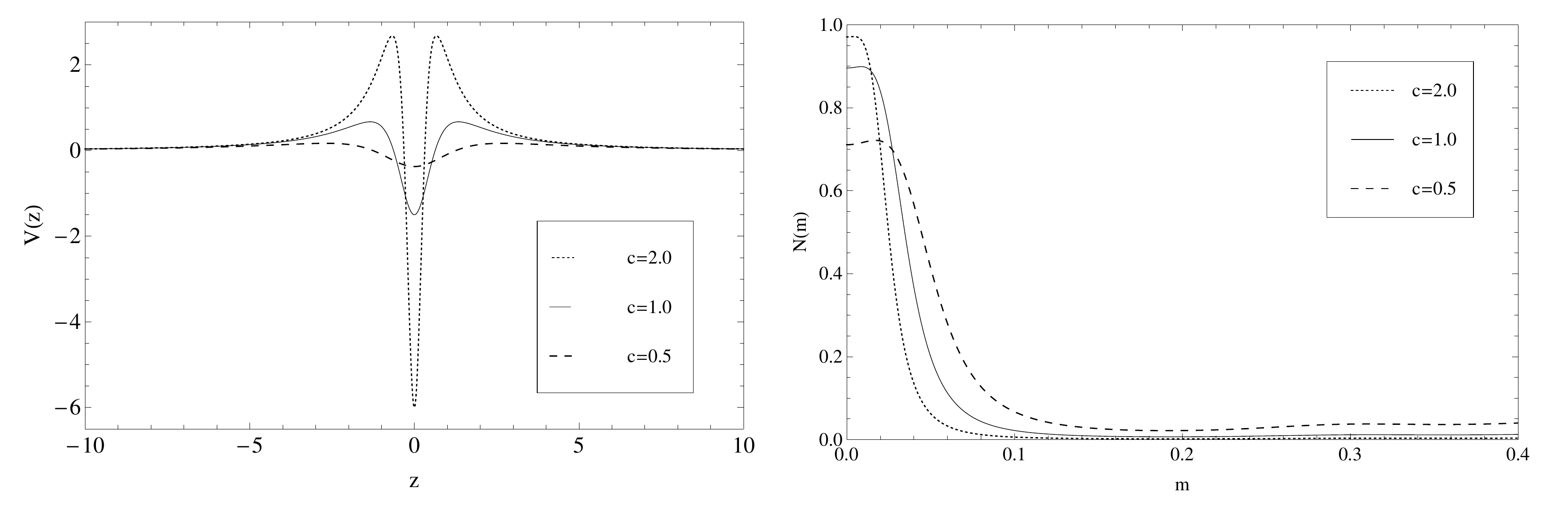}
 \caption{Schr\"{o}dinger potential (left) and the $N(m)$ function (right) for SG model.}
 \label{fig:sine1_1}
\end{figure*}

Now, we focus attention on the resonant modes. With the above solution to $A(z)$ we find the following potential
\begin{equation}\label{pot_sg}
V(z)=\frac{3c^2(-2+5c^2z^2)}{4(1+c^2z^2)^2}.
\end{equation}
This result allows us to determine the relative probability function $N(m)$, and we depict it in Fig. (\ref{fig:sine1_1}), together with the potential $V(z)$.
We note the presence of a localized graviton mode with $m=0$, which is a bound mode but not a resonance state. Besides, there is no peak in $N(m)$ for $m\neq0$. The absence of massive resonances can be understood in terms of the potential structure. The mass interval where the resonances could appear ($m^2\leq V_{max}$) is very thin due to the maxima of the volcano-like potential, that is also very low as showed in Fig. (\ref{fig:sine1_1}). Increasing the value of the constant $c$ the maxima of the potential are raised. However, its two barriers become very close so that the modes are suppressed at the brane location.

\subsection{DSG brane}

\begin{figure*}
 \centering
    \includegraphics[width=0.95\textwidth]{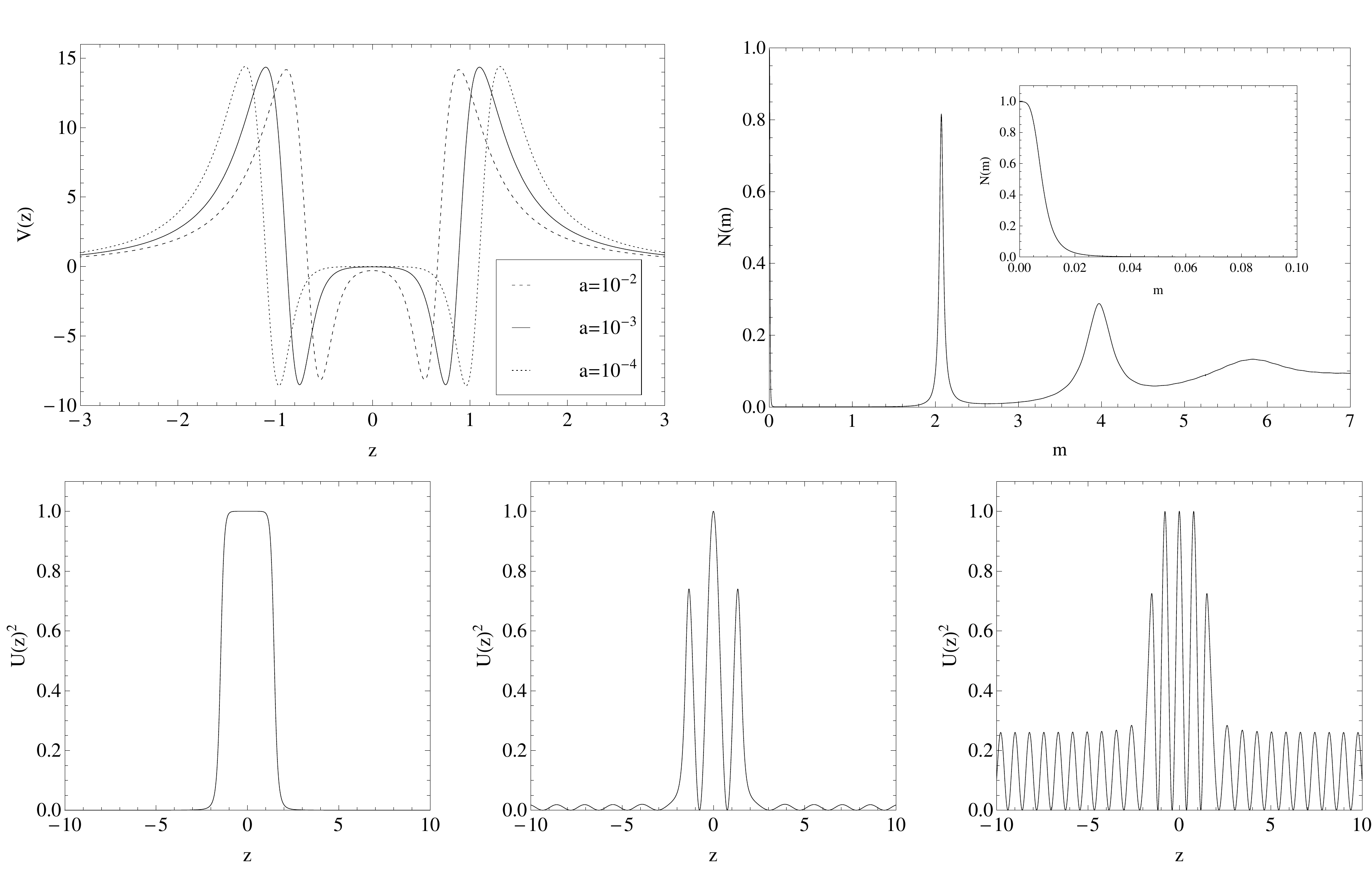}
 \caption{The Schr\"{o}dinger potential (top left), N(m) function (top right), zero mode (bottom left), $m=2.07$ resonance (bottom center) and $m=3.99$ resonance (bottom right). DSG case.}
 \label{fig:sine5}
\end{figure*}

To study the massive spectrum for the DSG model, we need to solve again the Schr\"{o}dinger equation (\ref{eq:schro}) and construct the $N(m)$ function (\ref{rel_prob}) in terms of the action (\ref{eq:action_dsg}). Given $\mathcal{A}(y)$ from Eq. (\ref{eq:eqA}) we perform the transformation $dz=e^{-A(y)}dy$ to get $\mathcal{A}(z)$ and the potential  $V(z)=\frac32\,\mathcal{A}^{\prime\prime}(z)+\frac94\,\mathcal{A}^{\prime2}(z)$. In Fig. (\ref{fig:sine5}) we plot the profile of the potential $V(z)$, for some values of $a$. 

The Schr\"{o}dinger potential to DSG model has two minima with a flat region at $z=0$. In contrast with the SG model, when we make the brane thicker (reducing $a$), the distance between the maxima of the $V(z)$ potential enlarge, but the height of the maxima do not reduce. The depth of the minima is also not reduced while increasing the thickness of the defect. This is an important feature since the resonant modes occurs for $m^2\leq V_{max}$. Thus, the Schr\"{o}dinger potential structure tell us that the DSG model is able to support resonant modes to the graviton.

In order to confirm the assumption of the existence of resonant modes, we obtain the relative probability $N(m)$ after integrating the Eq.(\ref{rel_prob}) numerically on the DSG setup. The result is showed in Fig. (\ref{fig:sine5}) using $a=10^{-6}$. We note the present of two massive resonances at $m=2.07$ and $m=3.99$. The zero mode is observable on the inset in Fig. (\ref{fig:sine5}) (top right)). Its structure is similar to the SG zero mode presented in Fig. (\ref{fig:sine1_1}). This result at $m=0$ indicates the existence of a graviton localized zero mode, which is confirmed by the solution of Eq. (\ref{eq:schro}) showed in Fig. (\ref{fig:sine5}) (bottom left).

\section{\label{conc}Conclusions}
In this work, we have considered two brane scenarios given by the sine-Gordon and the double sine-Gordon models. We focus our attention on the analysis of the graviton massive spectrum and resonances. 

Initially, the thick brane model generated by the SG potential was revised. The resulting solutions for the scalar field and the energy density show us that the thickness of the defect can be changed in terms of a parameter $c$ of the superpotential function. The curvature is positive at $y=0$ and becomes negative when $y\rightarrow\infty$. A Schr\"{o}dinger-like equation was obtained from the graviton equations of motion. Thereby, using the quantum mechanics interpretation, we have numerically evaluated the relative probability to find the modes on the brane location. We only found a peak in $N(m)$ at $m=0$, and it is correspondent to the zero mode localized.

The absence of massive resonances on the graviton spectrum of the SG model can be explained by the Schr\"{o}dinger potential structure of Eq. (\ref{pot_sg}). The maxima of such potential present small values and this excludes the possibility of massive resonances, that arise on the interval $m^2\leq V_{max}$. In fact, we can lift the maxima of the potential, but they became very narrow and again the possibility of resonance to the wave functions is excluded.

Changing our attention to the DSG brane model, we have noted the presence of a transient minima between the minima of the scalar field potential. Such behaviour is also expressed on the structure of solutions to the scalar field, as showed in Fig. \ref{fig:sine3}. Moreover, observing the solutions to the warp factor, energy density and scalar curvature, we have detected the existence of a flat region at the brane location. Another interesting detail of the DSG model is the presence of two maxima in $\mathcal{R}(y)$ and $\varepsilon(y)$  around the brane position. These new features indicate the possibility to capture modes inside the brane.

The Schr\"{o}dinger potential to the DSG model also presents significant differences from that of the SG model. The minimum point at $z=0$ splits in two, separated by a flat region where the potential assumes zero value. Moreover, when we make the brane thicker, the two maxima separate themselves without reduce their peaks. Therefore, there is a possibility that the wave functions acquire large amplitudes inside the brane. 

The existence of resonances were confirmed by the calculation of the relative probability function given by Eq. (\ref{rel_prob}), plotted in Fig. (\ref{fig:sine5}). We have detected two resonances respectively at $m=2.07$ and $m=3.99$. The peak in the function $N(m)$ corresponding to the zero mode is kept in accordance with the existence of a zero mode localized. However,  the appearance of massive resonances is related with the raising of the internal structure.

A natural extension of the current work consists in applying the same methodology to the other fields, including fermions and gauge fields, as in Ref. \cite{ca}, and investigate how the massive modes of those fields are trapped inside the SG and DSG branes. Another issue of interest is about the corrections to Newton's Law due to massive modes, where the dynamics of the eigenfunction $U_{m}(z)$ could modify the four-dimensional gravitational potential. Some of these issues are currently under investigation and will be reported in future work.



\section*{Acknowledgments}
This work was supported by the Brazilian
agencies Coordena\c{c}\~ao de Aperfei\c{c}oamento de Pessoal de
N\'{i}vel Superior (CAPES), the Conselho Nacional de Desenvolvimento
Cient\'{i}fico e Tecnol\'ogico (CNPq), and Fundacao Cearense de apoio ao Desenvolvimento Cientifico e Tecnologico (FUNCAP). The authors thank the anonymous referee for comments and suggestions.


\end{document}